\documentclass[aps,pre,twocolumn,showpacs,superscriptaddress,floatfix]{revtex4}
\usepackage{amsmath,amsfonts,amssymb,graphicx}

\begin{document}

\title{A Paradoxical Property of the Monkey Book}

\author{Sebastian Bernhardsson}\email{sebbeb@tp.umu.se}
\affiliation{IceLab, Department of Physics, Ume{\aa} University, 901 87 Ume{\aa}, Sweden}
\author{Seung Ki Baek}
\affiliation{IceLab, Department of Physics, Ume{\aa} University, 901 87 Ume{\aa}, Sweden}
\author{Petter Minnhagen}
\affiliation{IceLab, Department of Physics, Ume{\aa} University, 901 87 Ume{\aa}, Sweden}

\date{\today}

\begin{abstract}
A ``monkey book'' is a book consisting of a random distribution of letters and blanks, where a group of letters surrounded by two blanks is defined as a word. We compare the statistics of the word distribution for a monkey book with the corresponding distribution for the general class of random books, where the latter are books for which the words are randomly distributed. It is shown that the word distribution statistics for the monkey book is different and quite distinct from a typical sampled book or real book. In particular the monkey book obeys Heaps' power law to an extraordinary good approximation, in contrast to the word distributions for sampled and real books, which deviate from Heaps' law in a characteristics way. The somewhat counter-intuitive conclusion is that a ``monkey book'' obeys Heaps' power law precisely because its word-frequency distribution is \emph{not} a smooth power law, contrary to the expectation based on simple mathematical arguments that if one is a power law, so is the other.
\end{abstract}


\maketitle

\subsection{Introduction}
Words in a book occur with different frequencies. Common words like ``the'' occur very frequently and constitute about 5\% of the total number of written words in the book, whereas about half the different words only occur a single time \cite{baayen_book}. The word-frequency $N(k)$ is defined as the number of words which occur $k$-times. The corresponding word-frequency distribution (wfd) is defined as $P(k)=N(k)/N$ where $N$ is the total number of different words. Such a distribution is typically broad and is often called ``fat-tailed''  and ``power law``-like. ``Power law''-like means that the large $k$-tail of the distribution to a reasonable approximation follows a power law, so that $P(k)\propto 1/k^\gamma$. Typically, one finds that $\gamma\leq 2$ for a real book \cite{seb_1,seb_2,newman,Zipf_1,Zipf_2,Zipf_3}. What does this broad frequency distribution imply? Has it something to do with how the book is actually written? Or has it something to do with the evolution of the language itself? The fact that the wfd has a particular form was first associated with the empirical Zipf-law for the corresponding word-rank distribution.\cite{Zipf_1,Zipf_2,Zipf_3} Zipf's law corresponds to $\gamma=2$. Subsequently Herbert Simon proposed that the particular form of the wfd could be associated with a growth model, the Simon model, where the distribution of words was related to a particular stochastical way of writing a text from the beginning to the end.\cite{simon} However, a closer scrutiny of the Simon model reveals that the statistical properties implied by this model are fundamentally different from what is found in any real text.\cite{seb_2} Mandelbrot (at about the same time as Simon suggested his growth model) instead proposed that the language itself had evolved so as to optimize an information measure based on an estimated word cost (the more letters needed to build up a word the higher cost for the word).\cite{mandelbrot,mitzenmacher} Thus in this case the power law of the word-distribution was proposed to be a reflection of an evolved property of the language itself. However, it was later pointed out by Miller in Ref.\ \cite{miller} that you do not need any particular language-evolution optimization to obtain a power law: A monkey randomly typing letters and blanks on a type-writer will also produce a wfd which is power-law like within a continuum approximation. The monkey book, hence, at least superficially have properties in common with real books \cite{Li,Ferrer09,Ferrer02,Ferrer10,Conrad04}. The case that the relation to real books are just superficial have in particular been argued in Refs \cite{Ferrer09} and \cite{Ferrer10}.

In 1978, Harold Stanley Heaps \cite{heap} presented another empirical law describing the relation between the number of different words, $N$, and the total number of words, ${M}$. Heaps' power law states that $N(M) \propto M^{\alpha}$, where $\alpha$ is a constant between zero and one. However, it was recently shown that Heaps' law gives an inadequate description of this relation for real books, and that it needs to be modified so that the exponent $\alpha$ changes with the size of the book from $\alpha=1$ for $M=1$ to $\alpha=0$ as $M \rightarrow \infty$ \cite{seb_1}. 
It was also shown that the wfd of real books, in general, can be better described by introducing an exponential cut off so that $P(K)=A\exp(-bk)k^{-\gamma}$ \cite{seb_2}.
A simple mathematical derivation of the relation between the power-law exponents $\gamma$ and $\alpha$ gives the result $\alpha = \gamma - 1$ \cite{seb_1}. This in turn means that the shape of the wfd also changes with the size of the book, so that $\gamma=2$ for small $M$, but reaches the limit value $\gamma=1$ as $M$ goes to infinity. The same analysis showed that the parameter $b$ is size dependent according to $b \approx b_0/M$ \cite{seb_1}.
It was also shown empirically that the works of a single author follows the same $N(M)$-curve to a good approximation
and which was further manifested  in the \emph{meta-book concept}: the $N(M)$-curve  characterizing  a text of an individual author  is obtainable by pulling sections from  the author´s collective meta book.\cite{seb_1}  As will be further discussed  below, the shape of the $N(M)$ curve is mathematically closely related to the \emph{Random Book Transformation} (RBT) \cite{seb_1}\cite{seb_2}.

As mentioned above, the writing of a real book cannot be described by a growth model because the statistical properties of a real book are translational invariant \cite{seb_2}. The monkey book, on the other hand, is produced by a translational-invariant stationary process.  An important question of much attention is then how close the statistical properties of the monkey book really are to those of a real book. It is shown in the present work that in the context of Heaps' law the answer is somewhat paradoxical.

\subsection{Monkey book}
Imagine an alphabet with $\mathcal{A}$-letters and a typewriter with a keyboard with one key for each letter and a space bar. For a monkey randomly typing on the typewriter the chance for hitting the space bar is assumed to be $q_s$ and the chance for hitting any of the letters is $\frac{(1-q_s)}{\mathcal{A}}$. A word is then defined as a sequence of letters surrounded by blanks. What is the resulting wfd for a text containing ${M}$ words? Miller in Ref.\ \cite{miller} found that in the continuum limit this is in fact a power law. 
In the appendix we re-derive this result using an information cost method. A more standard alternative derivation can be found in Ref.\ \cite{newman}.

We will denote the word-frequency distributions in the continuum limit by $p(k)$ and in the Monkey book case it is given by

\begin{equation}
p(k) \propto \frac{1}{k^{\gamma}}
\end{equation}

with

\begin{equation}
\gamma=\frac{2\ln \mathcal{A}-\ln (1-q_{s})}{\ln \mathcal{A}-\ln (1-q_{s})}
\label{gamma}
\end{equation}

Thus, if $q_s = 1/(\mathcal{A}+1)$ then $\gamma=1$ if $\mathcal{A}=1$ and $\gamma = 2$ in the infinite limit of $\mathcal{A}$.

\subsubsection{Continuum approximation versus real word-frequency\\}

\begin{figure}
\centering
\includegraphics[width=\columnwidth]{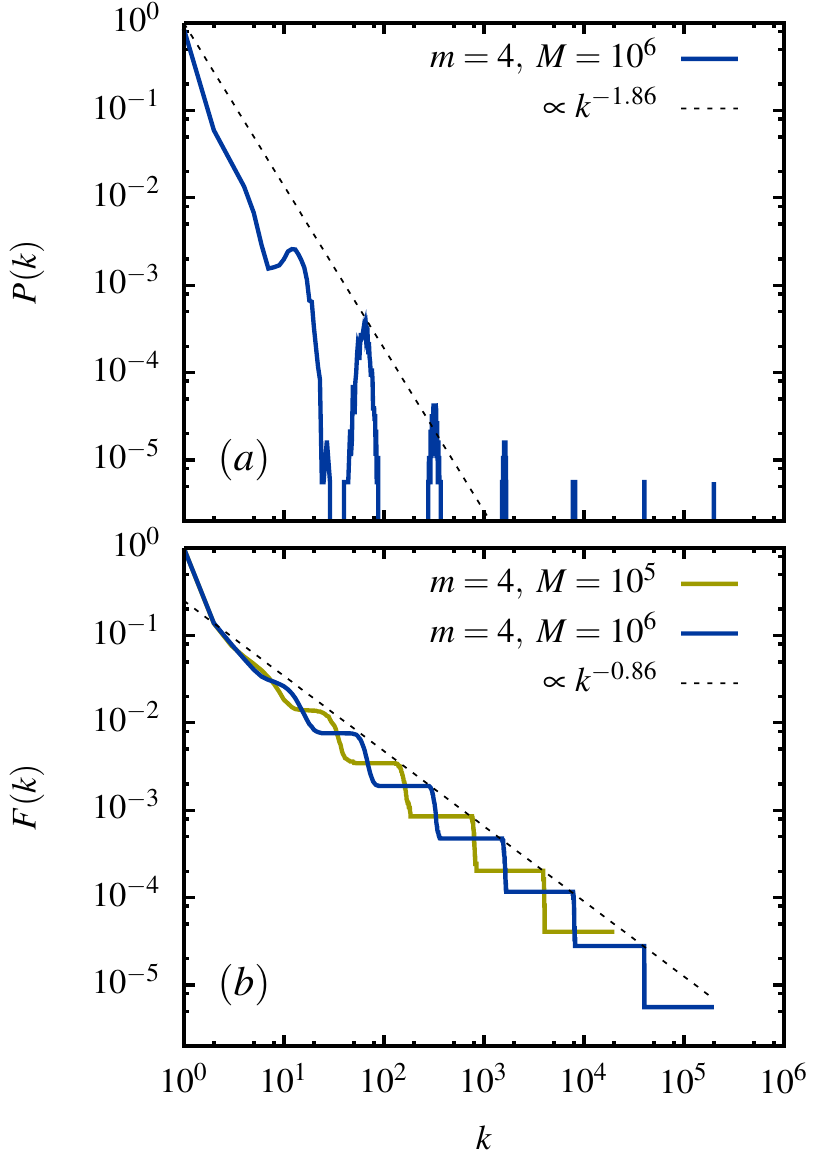}
\caption{Word-frequency distribution for the monkey book. (a) Broken straight line corresponds to the continuum approximation $p(k)\propto k^{-\gamma}$ given by Eq.\ \ref{gamma}, whereas the full curve with disjunct peaks represents the real distribution $P(k)$. $P(k)$ and its continuum approximation $p(k)$ are clearly very different.(b) The corresponding cumulative distributions $f(k)$ and $F(k)$. Broken straight line corresponds to $f(k)\propto k^{-(\gamma-1)}$ and the black zig-zag line to the corresponding real cumulative distribution $F(k)$. Note that $f(k)$ to good approximation is an envelope of the black zig-zag $F(k)$. The gray zig-zag-curve is the cumulative $F(k)$ for a tenth of the monkey book. Note that $f(k)$ still gives an equally good envelop. Thus the envelop of the cumulative $F(k)$ for a monkey book is a \emph{size-independent} power law.}
\label{fig1}
\end{figure}

The above result for $p(k)$ is an approximation of the actual (discrete) result expected from random typing. The true wfd of the model will here be denoted as $P(k)$. What is then the relation between the power-law form of $p(k)$ and the actual probability, $P(k)$, for a word to occur $k$-times in the text? It is quite straight-forward to let a computer take the place of a monkey and simulate monkey books \cite{Li}. Fig.\ 1a gives an example for an alphabet with $\mathcal{A}=4$ letters, a total lumber of words ${M}=10^6$ and with the chance to hit the space bar $q_s=1/(\mathcal{A}+1)=1/5$. Such a book should have a power-law exponent of $\gamma\approx1.86$ according to Eq.\ \ref{gamma}.
Note that $P(k)$ for higher $k$ consists of disjunct peaks: the peak with the highest $k$ corresponds to the $\mathcal{A}=4$ one-letter words, the next towards lower $k$ to the $\mathcal{A}^2=16$ two-letter words and so forth. Thus the power law tail $1/k^\gamma$ in the case of a monkey-book is not a smooth tail but a sequence of separated peaks as previously reported in Ref.\ \cite{Li,Ferrer09,Ferrer02,Ferrer10,Conrad04}. So what is the relation to the continuum $p(k) \propto k^{-1.86}$? Plotted in log-log scales as in Fig.\ 1a, $p(k)$ is just a straight-line with the slope $-\gamma=-1.86$ (broken line in Fig.\ 1a). Represented in this way there is no obvious discernible relation between the separated peaks of $P(k)$ and the straight line given by $p(k)$. In order to directly see the connection one can instead compare the cumulative distributions $F(k)=\sum_{k'=k}^{M} P(k')$ and $f(k)=\sum_{k'=k}^{M} p(k')\propto 1/k^{0.86}$. In Fig.\ 1b, $F(k)$ corresponds to the full drawn zig-zag-curve and the straight broken line with slope $-0.86$ to the continuum approximation $f(k)$. In this plot the connection is more obvious: $f(k)$ is an envelope of $F(k)$. 
Figure 1b also illustrates that the envelop slope for the monkey book is independent of the length of the book: The full drawn zigzag curve corresponds to ${M}=10^6$ whereas the dotted zigzag curve corresponds to ${M}=10^5$. Both of them have the envelop slope $-\gamma=-0.86$ given by the continuum approximation $f(k)$. 

\emph{To sum up}: The continuum approximation $p(k)\propto 1/k^\gamma$ is very different from the actual spiked monkey book, $P(k)$.
However, the envelop, $f(k)$, for the cumulative wfd, $F(k)$, of the monkey book is nevertheless a power law with a slope which is independent of the size of the book.

\subsection{Heaps' law}
Heaps' law is an empirical law which states that the number of different words, $N$, in a book approximately increases as $N(M)\propto {M}^\alpha$ as a function of the total number of words \cite{heap}. For a random book, like the monkey book, there is a direct connection between $P(k)$ and the $N(M)$-curve. A random book means a book where the class of words which occurs $k$ times are randomly distributed throughout the book: The chance of finding a word with frequency $k$ is independent of the position in the book i.e. it is as likely to find a word with a frequency $k$ at the beginning, in the middle or at the end of the book. Suppose that such a book of size $M$ has a wfd $P_M(k)$ created by sampling a fixed theoretical probability distribution $p(k)\propto k^{-\gamma}$, where the normalization constant is only weakly dependent on ${M}$. The number of different words for a given size is then related to ${M}$ through the relation

\begin{equation}
{M} =N({M})\sum_{k=1}^{M} kp(k)
\label{heap}
\end{equation}

 and, since in the present case  

\begin{equation}
 \sum_{k=1}^{M} kp(k) \propto \frac{1}{2-\gamma}({M}^{2-\gamma}-1),
 \label{power_heap}
\end{equation}

it follows that 

\begin{equation}
N(M)\propto {M}^{\gamma-1}.
\label{heaps_law}
\end{equation}

A heuristic direct way to this result is to argue that the first time for a word with frequency $k$ to occur is inversely proportional to its frequency $\tau\propto 1/k$, so that you in the time-interval $[\tau,\tau+d\tau]$ introduce $n(\tau)d\tau\propto \frac{1}{k^\gamma}|\frac{dk}{d\tau}|d\tau\propto\tau^\gamma\tau^{-2} d\tau$ new words. Since $\tau$ is proportional to how far into the book you are, this means that ${N}\propto \int_0^{M}\tau^{\gamma-2}d\tau\propto {M}^{\gamma-1}$. The conclusion from Eqs.\ \ref{heap}-\ref{heaps_law} is that the $N(M)$-curve of a random book with $P_M(k)\propto k^{-\gamma}$ should follow Heaps' law very precisely with the power-law index $\alpha=\gamma-1$. One consequence of this is that if you start with such a book of size ${M}$ and the number of different words $N({M})$ and then randomly pick half the words, then this new book of ${M}/2$ words will on the average have $N({M}/2) \propto ({M}/2)^{\alpha=\gamma-1}$ different words. Thus, starting from a random book of size ${M}$, you can obtain the complete $N({M})$-curve by dividing the book into parts of smaller sizes. Furthermore, in the special case where $P_M(k)$ is a power law with a functional form, and a power-law index, which is \emph{size-independent}, the $N({M})$-curve follows Heaps' law very precisely with $\alpha=\gamma-1$. 
Figure 2 illustrates that this is indeed true for monkey books by showing the $N(M)$-curve for different alphabet sizes (full drawn curves) together with the corresponding analytic solutions (broken curves). Note that for Heaps' law, $N(M) \propto M^{\alpha}$, and the relationship $\alpha=\gamma-1$ to hold, the full curves should be parallel to the broken curves for each alphabet size, respectively. Also, the continuum theory from Eq.\ \ref{gamma} gives $\gamma=1$ for $\mathcal{A}=1$ (an alphabet with a single letter) which by Eq.\ \ref{heaps_law} predicts $N\propto \ln M$, and which is again in full agreement with the monkey book.

\begin{figure}
\centering
\includegraphics[width=\columnwidth]{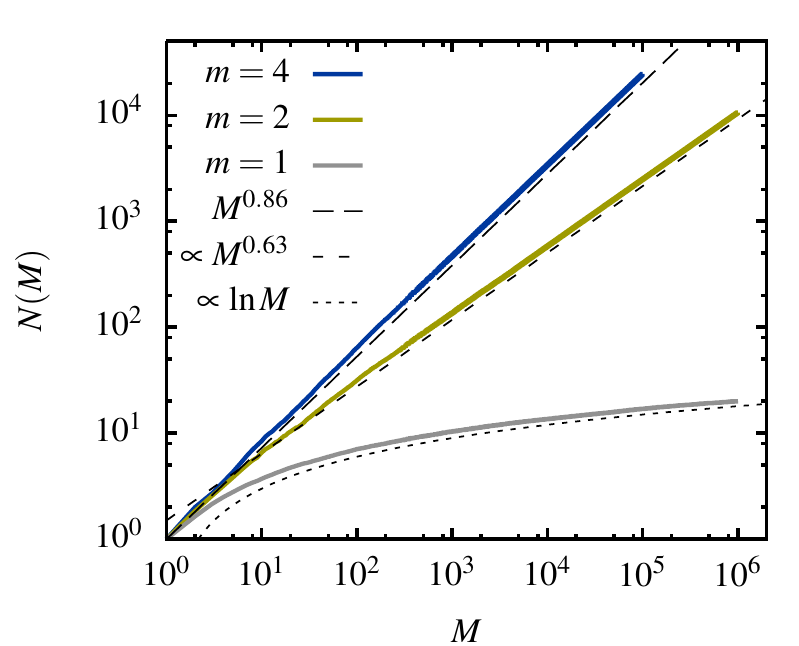}
\caption{Heaps law for monkey books with different sizes of the alphabet, in log-log scale. The full curves from top to bottom gives the $N(M)$ for alphabets of length $m=4, 2,$ and $1$, respectively. According to Eq.(7) the $N(M)$ should for $m=4$, and $2$ follows Heaps power laws with the exponents $0.86$ and $0.63$, respectively, and the corresponding broken lines show that these predictions are borne out to excellent precision. For $m=1$, Eq.(7) predicts that $N(M)$ instead should be proportional to $\ln M$, since $\gamma-1=0$. The corresponding broken curve again shows an excellent agreement. }
\label{fig2}
\end{figure}

However, notwithstanding this excellent agreement, the reasoning is nevertheless flawed by a serious inconsistency:
The connection to Heaps' law, $N(M) \propto M^{\alpha}$, was here established for a random book with a continuous power-law wfd, whereas the wfd of a monkey book consists of a series of disjunct peaks, and only the envelope of its cumulative wfd can be described by a continuous power law. It thus seems reasonable that a random book with a wfd which is well described by a smooth power law would satisfy Heaps' law to an even greater extent. However, this reasoning is not correct. The derived form of Heaps' law, $N(M) \propto M^{\gamma-1}$ is based on a wfd for which the functional form and $\gamma$ is \emph{size independent}. 
But as we will show in the following section, this is an impossibility: a random book with a continuous wfd can in principle \emph{not} be described by a size-independent power-law.

\subsection{Contradicting power laws}

The most direct way to realize this inconsistency problem is to start from a random book which has a smooth power-law wfd with an index $\gamma$.
Such a book can be obtained by randomly sampling word frequencies from a continuous power-law distribution of a given $\gamma$ and then placing them, separated by blanks, randomly on a line. For this ''sampled book'' one can then directly obtain the $N({M})$-curve by dividing the book into parts, as described above. Fig.\ 3a gives an example of a $N(M)$-curve for a sampled book with $\gamma=1.86$, $N=10^5$ and ${M}=10^6$. The resulting wfd is shown in Fig.\ 3b.

It is immediately clear from Fig.\ 3a that a sampled book with a power-law wfd does \emph{not} have an $N({M})$-curve which follows Heaps' law, $N(M) \propto M^{\alpha}$ (it deviates from the straight line in the figure). This is thus in contrast to the result of the derivation given by Eq.\ \ref{heap}-\ref{heaps_law} and the monkey-book which does obey Heaps' law, $N(M) \propto M^{\gamma-1}$, as seen from Fig.\ 2. This means that the monkey book obeys Heaps' law \emph{because} the wfd is \emph{not} well described by a smooth power law, and that the ''spiked'' form of the monkey-$P(k)$ is, in fact, crucial for the result.
The explanation for the size invariance of the monkey book can be found in the derivation presented in the appendix. Since the frequency of each word is exponential in the length of the word, it naturally introduces a discrete size-invariant property of the book. This discreteness is responsible for the disjunct peaks shown in Fig.\ 1a, and it is easy to realize that non-overlapping Gaussian peaks will transform into new Gaussian peaks with conserved relative amplitudes, thus resulting in a size-independent envelope.

\begin{figure}
\centering
\includegraphics[width=\columnwidth]{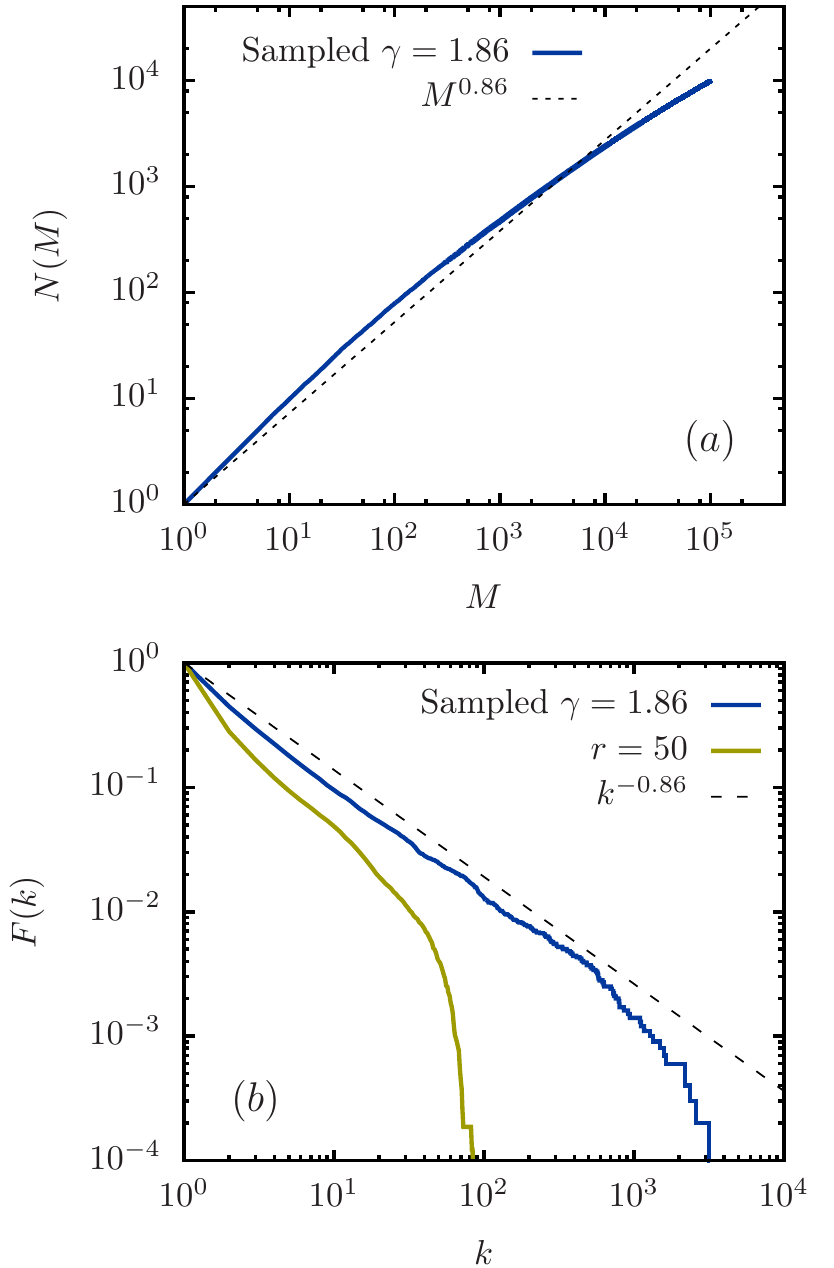}
\caption{Results for a ``sampled book'' of length $M$ described by a smooth power law wfd $P(k)\propto k^{-\gamma}$. (a) Full drawn curve is the real $N(M)$ whereas the broken straight line is the Heaps` power law prediction from Eq.(7). Since the real $N(M)$-curve is bent, it is clear that a power law wfd does not give a power law $N(M)$. 
(b) illustrates that the wfd obtained for a part of the full book containing $M'$ words where $r=M/M'$ has a different functional form than $P_{M}$. The curves show the cumulative distributions $F(k)=\sum_{k'=k}^{M} P(k')$ for the full random book $M=10^6$ and  $M'=5000$, respectively.}
\label{fig3}
\end{figure}

The core of this paradoxical behavior lies in the fact that the derived form of the $N(M)$-curve requires a size-independent wfd, and that a random book is always subject to well-defined statistical properties. One of these properties is that the $P_{M}(k)$ transforms according to the RBT (random book transformation) when dividing it into parts \cite{seb_1,baayen_book}:
The probability for a word that appears $k^{\prime}$ times in the full book of size ${M}$ to appear $k$ times in a smaller section of size ${M}'$ can be expressed in binomial coefficients: Let $P_{M}(k^{\prime})$ and $P_{{M}'}(k)$ be two column matrices with elements numerated by $k^\prime$ and $k$, then

\begin{equation}
\boldsymbol{P}_{{M}'}(k)=C\sum_{k^{\prime}=k}^{{M}}\boldsymbol{A}_{kk^{\prime}}\boldsymbol{P}_{{M}}(k^{\prime})
\label{1}
\end{equation}

where $A_{kk^{\prime}}$ is the triangular matrix with the elements%

\begin{equation}
A_{kk^{\prime}}=(r-1)^{k^{\prime}-k}\frac{1}{r^{k^{\prime}}}\Bigg( \begin{array}{c}k' \\k\end{array} \Bigg)
\label{2}
\end{equation}

and $r=M/M'$ is the ratio of the book sizes. The normalization factor $C$ is

\begin{equation}
C=\frac{1}{1-\sum_{k^{\prime}=1}(\frac{{M}-{M}'}{{M}})^{k^{\prime}}P_{M}(k^{\prime})}
\label{3}
\end{equation}

Suppose that $P_{M}(k)$ is a power law with an index $\gamma$. The requirement for the corresponding random book to obey Heaps' law is then that $P_{M}(k)$ under the RBT-transformation remains a power law with the same index $\gamma$. However, the RBT-transformation does not leave invariant a power law with an index $\gamma>1$ \cite{seb_1,seb_2}. This fact is illustrated in Fig.\ 3c, which shows that a power law $P_{M}(k)$ changes its functional form when describing a smaller part of the book. This change of the functional form is the reason for why the $N({M})$-curve in Fig.\ 3a does not obey Heaps' law. The implication of this is that a random book which is well described by the continuum approximation $p(k) \propto 1/k^{\gamma}$ can never have a $N(M)$-curve of the Heaps’ law form $N(M) \propto N^{\alpha}$.

In Fig.\ 4a-c we compare the result for a power law $P_{M}(k)$ in Fig.\ 3a-c to the real book \emph{Moby Dick} by Herman Melville. Fig.\ 4a shows the $N({M})$ for ${M}\approx 212000$ both for the real book and for the randomized version (where the words in the real book are randomly re-distributed throughout the book) \cite{seb_2}. 
As seen, the $N(M)$-curve for the real and randomized book are closely the same and very reminiscent of the pure power-law case in Fig.\ 3a: A real and random book, as well as a power-law book, deviates from Heaps' law in the same way. In Fig.\ 4c we show that the reason is the same: The form of the wfd changes with the size of the book in similar ways. The result for the real book is not a property solely found in Moby Dick, but has previously been shown to be an ubiquitous feature of novels \cite{seb_1}.

\begin{figure}
\centering
\includegraphics[width=\columnwidth]{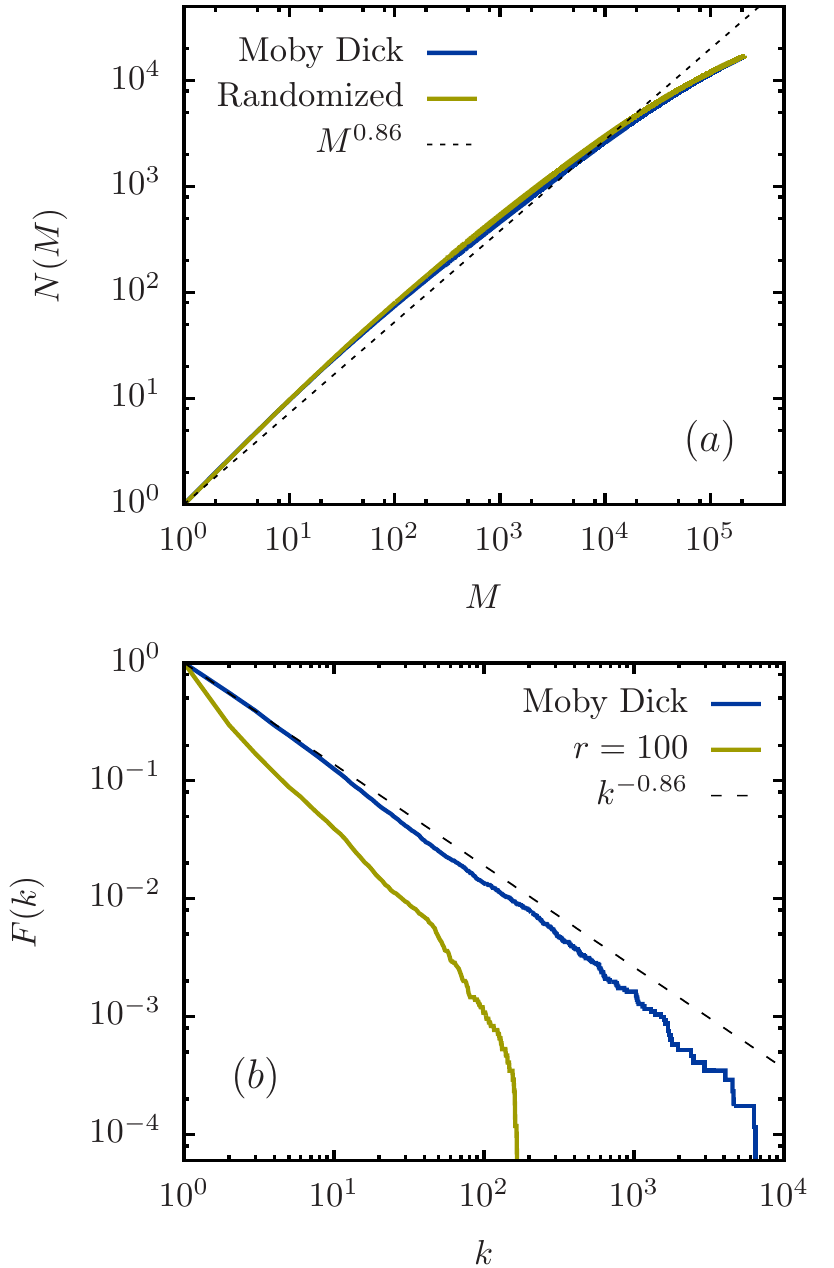}
\caption{Comparison with a real book. (a) $N(M)$-curves for \emph{Moby Dick} (dark curve) and for the randomized \emph{Moby Dick} (light curve) together with a power law (straight broken line). Real and random \emph{Moby Dick} has to excellent approximation the same $N(M)$ and this $N(M)$-curve is not a power law. Note the striking similarity with Fig.3a. 
(b) Change in the cumulative distribution $F(k)$ with text length for \emph{Moby Dick}, dark curve corresponds to the full length $M_{tot}\approx 212000$ words and the light curve to $M'\approx 2000$ ($r=M/M' = 100$). The change in the functional form of the wfd is very similar to the power law book shown in Fig.3b.}
\label{fig4}
\end{figure}

\emph{To sum up:}
A simple mathematical derivation tells us that if the wfd is well described by a power law, then so is the $N(M)$-curve. This power-law form $N(M) \propto M^{\alpha}$ is called Heaps' law.
However, a sampled book, as well as real books, does not follow Heaps' law, in spite of the fact that their wfds are well described by smooth power laws. In contrast, the monkey book which has a spiky, disjunct, wfd, does obey Heap's law very well.

\subsection{Conclusions}
We have shown that the $N(M)$-curve for a monkey book obeys Heaps' power-law form $N(M)\propto M^{\alpha}$ very precisely. This is in contrast to  real and randomized real books, as well as sampled books with word-frequency distributions (wfd) which are well described by smooth power laws: All of these have $N(M)$-curves which deviate from Heaps' law in similar ways.
In addition we discussed the incompatibility of simultaneous power-law forms of the wfd and the $N(M)$-curves (Heaps' law). This led to the somewhat counter-intuitive conclusion that Heaps' power law requires a wfd which is \emph{not} a smooth power law!
We have argued that the reason for this inconsistency is that the simple derivation that leads to Heaps' law when starting from a power-law wfd assumes that the functional form is size independent when sectioning down the book to smaller sizes. However, it is shown, using the Random book transformation (RBT), that this assumption is in fact not true for real or randomized books, nor for a sampled power-law book. In contrast, a monkey book, which has a spiked and disjunct wfd, possesses an invariance under this transformation.
It is shown that this invariance is a direct consequence of the discreteness in the frequencies of words due to the discreteness in the length of the words (see appendix).

\section{Appendix: The information cost method}

Lets imagine a monkey typing on a keyboard with $\mathcal{A}$ letters and a space bar, where the chance for typing space is $q_s$ and for any of the letters  is $\frac{(1-q_s)}{\mathcal{A}}$.
A text produced by this monkey has a certain information content given by the entropy of the letter configurations produced by the monkey.
These configurations result in a word frequency distribution (wfd) $P(k)$ and the corresponding entropy $S=-\sum_k P(k)\ln P(k)$ gives a measure of the information associated with this frequency distribution. The most likely $P(k)$ corresponds to the maximum of $S$ under the appropriate constraints. This can equivalent be viewed as the minimum information loss, or cost, in comparison with an unconstrained $P(k)$ \cite{info_book}.
Consequently, the minimum-cost $P(k)$ gives the most likely wfd for a monkey.

Since the wfd in the continuum approximation is different from the real distribution $P(k)$, we will call the former $p(k)$. 
Let $k$ be the frequency with which a specific word occurs in a text and let the corresponding probability distribution be $p(k)dk$. This means that $p(k)dk$ is the probability that a word belongs to the frequency interval $[k,k+dk]$. The entropy associated with the probability distribution $p(k)$ is $S=-\sum_{k}p(k)\ln p(k)$ (where $\sum_{k}$ implies an integral whenever the index is a continuous variable). Let ${M}(l)dl$ be the number of words in the word-letter length interval $[l,l+dl]$. This means that the number of words in the frequency interval $[k,k+dk]$ is ${M}(l)\frac{dl}{dk}dk$ because all words of a given length $l$ occur with the same frequency. The number of distinct words in the same interval is $n(k)dk=Np(k)dk$, which means that $\frac{{M}(l)}{n(k)}\frac{dl}{dk}$ is the degeneracy of a word with frequency $k$. The information loss due to this degeneracy is $\ln(\frac{{M}(l)}{n(k)}\frac{dl}{dk})=\ln({M}(l)\frac{dl}{dk})-\ln p(k)+const$(in nats). The average information loss is given by       
\begin{equation}
I_{cost}=\sum p(k)[-\ln p(k)+\ln ({M}(l)dl/dk)]
\label{icost}
\end{equation}
and this is the appropriate information cost associated with the words: The $p(k)$ which minimizes this cost corresponds to the most likely $p(k)$.
The next step is to express ${M}(l)$ and $dl/dk$ in terms of the two basic probability distributions, $p(k)$ and the probability for hitting the keys: ${M}(l)$ is just ${M}(l)\sim \mathcal{A}^{l}$. The frequency $k$ for a world containing $l$ letters is

\begin{equation}
k\sim (\frac{1-q_{s}}{\mathcal{A}})^{l}q_{s}
\label{kl}
\end{equation}

Thus $k\sim \exp (al)$ with $a=\ln (1-q_{s})-\ln \mathcal{A}$ so that $dk/dl=ka$ and, consequently, $I_{loss}=-\sum p(k)\ln p(k)+\sum p(k)[\ln \mathcal{A}^{l}-\ln ka]$. Furthermore,
$\ln (\mathcal{A}^{l}/ka)=l\ln \mathcal{A}-\ln k-\ln a$ and from Eq.\ref{kl} one gets $l=\ln (k/q_{s})/\ln (1-q_{s})/\mathcal{A})$ from which follows that $\ln (\mathcal{A}^{l}/ka)=(-1+\frac{\ln \mathcal{A}}{\ln (1-q_{s})-\ln \mathcal{A}})\ln k+const$. Thus the most likely distribution $p(k)$ corresponds to the minimum of the information word cost  
\begin{equation}
I_{cost}=-\sum p(k)\ln p(k)+\sum p(k)\ln k^{-\gamma}
\end{equation}
with 
\begin{equation}
\gamma=\frac{2\ln \mathcal{A}-\ln (1-q_{s})}{\ln \mathcal{A}-\ln (1-q_{s})}
\label{gamma_appendix}
\end{equation}
Variational calculus then gives $\ln (p(k)k^{\gamma})=const$ so that
\begin{equation}
p(k)\propto k^{-\gamma}.
\end{equation}
Note that the total number of words ${M}$ only enter this estimate though the normalization condition. This means that the continuum approximation $p(k)\propto \frac{1}{k^{\gamma}}$ for the monkey-book is independent of how many words $M$ it contains. Thus if you start from a monkey-book with $M$ words and you randomly pick a fraction of these $M$ words, then this smaller book will also a have a wfd which in the continuum limit follows the same power-law. This is a consequence of the fact that the frequency $k$ for a word a length $l$ is always given by eq.(\ref{kl}) irrespective of the book-size. It is this specific monkey-book constraint which makes $I_{cost}$ in eq.(\ref{icost}) $M$-invariant and hence forces the continuum $p(k)$ to always follow the same power-law. The crucial point to realize is that the very same constaint forces the real $P(k)$ to have a "peaky" structure. One should also note that if you started from a book consisting of $M$ words randomly drawn from the \emph{continuum} $p(k)$ then a randomly drawn fraction from this book will no longer follow the original power-law.

\end{document}